# Astronomical Research with liquid mirror telescopes

Ermanno F. Borra,

Centre d'Optique, Photonique et Laser, Observatoire du Mont Mégantic,
Département de Physique, Université Laval, Canada G1K 7P4
http://wood.phy.ulaval.ca/lmt/home.html
Fax: 418-656-2040: Email: borra@phy.ulaval.ca

## ABSTRACT

I discuss the modalities of zenith observing with LMTs, their limitations and advantages. The main limitation comes from the relatively small regions of the sky that LMTs can access. The main advantage comes from low cost, for inexpensive large LMTs can be dedicated to very specialized goals. Large surveys, cosmological studies, variability studies are prime examples of astronomical research where LMTs can have a major impact. I examine the performance of a 4-m telescope equipped with wide band filters and its use for astronomical research. The data could be used to select targets for the VLT. I examine the usage of the data for variability studies and, in particular, for a supernova search.

## 1. INTRODUCTION

Liquid mirrors are a new technology that has begun to demonstrate a performance that is interesting to astronomy. As discussed in my earlier presentation , and by others, in these proceedings, liquid mirrors have been demonstrated in the laboratory (up to 2.5-m diameters) and on the sky (up to 3-m diameters). There is therefore a window of opportunity in which one can do what has never been done before in Astronomy: dedicate a 4-m class telescope to a full time project. This is rendered possible by the low cost of LMTs. A fully instrumented 4-m class LMT observatory should cost of the order of US$500,000 ( - $0.0 + US$ 200,000). Liquid mirrors cost almost two orders of magnitude less than glass mirrors. Let us forget for a brief time that we are dealing with a LM and let us imagine that we have found a "surplus" 4-m mirror for a few tens of thousands of dollars. The challenge then is: What will we do with it, considering that we are unlikely to get more than a few $10^5$ dollars to build the telescope and observatory, instrument it and run it for a couple of years? This limits us to a transit telescope that will give the type of data one gets with a Schmidt telescope. We therefore get a 4-m Schmidt with no pointing capabilities.

The optimum size for a low-risk endeavor should be in the 4-m class, because it is a small extrapolation from the telescopes that have been built (Borra, these proceedings). Larger sizes would entail some R&D that we do not want to get into. The cost is also small enough that granting agencies should be willing to take the risk in this era of tight budgets. This size is also the size of the astronomical "workhorses" of the end of this century. It is hard to get more than a few nights a year on these telescopes: We can thus intuitively see that such a telescope working full time for a narrowly defined project should be able to do frontier research.

In this article, I will consider the data and the science that one can do with a LMT in the 4-m class. The survey should have a main science driver, which will dictate the telescope design and its instrumentation, with secondary projects. Obviously the driver should involve frontier astronomy.

## 2. OBSERVING WITH LMTs

LMTs can only observe the zenith. Observing with a zenith telescope is obviously quite different from observing with a tiltable telescope that can point and track. A LMT continuously monitors a strip of sky passing through the zenith so that the integration times are limited by the time it takes an object to cross the detector. It must be appreciated that, if on the one hand, TDI zenith observing has limitations, it also has a number of advantages with respect to conventional observing. Flatfielding and defringing are much more accurate because the images are actually formed by averaging over entire CCD columns (in the direction of the scan). Extinction and image size are optimized at the zenith. Observing efficiency is high because there is no overhead from slewing, reading out the CCD, taking flatfielding frames, etc.

### 2.1 Integration times and limiting magnitudes

TDI tracking with a zenith telescope restricts one to short integration times: the time it takes an object to drift across the detector. The nightly single-pass integration time is given by

$$t = 1.37 \ 10^{-2} \ n \ w \ /(f \cos (\text{lat})), \qquad (1)$$

where the time is in seconds, n is the number of pixels along the read out direction of the CCD, w the pixel width (microns), f the focal length of the telescope (meters) and lat is the latitude of the observatory.

Let us consider a 3.6-m diameter f/1.4 LMT equipped with a 2048X2048 CCD having 15 micron pixels ( 0.62 arcsec); it yields an integration time of 100 seconds/night. We can reach R = 23 for stellar objects with a total S/N = 5 and R = 22.2 with S/N = 10 (see Borra 1995 for the computations used ). Assuming 60 nights of observation per year, excluding thus moonlit time and including only the nights during which an object would actually cross the zenith, the limiting magnitudes would increase by 2.2. Lengthening the focal length to give 0.41 arcsec pixels improves the resolution but decreases the integration times, reducing slightly the limiting magnitude. This information is summarized in Table 1. Using a larger CCD, or a mosaic, increases the exposure times but increases costs. However, within a limited budget, the money is probably better spent by using a dichroic beam splitter and obtaining simultaneous color information with 2 or more CCDs.

Table 1
Limiting magnitudes  (2048x2048  CCD with 15 micron pixels)
=========================================================

| 0.62 Arcsec pixels, S/N=5 | | | | | |
| --- | --- | --- | --- | --- | --- |
| 1 night (100 sec) | | | 1 year (6000 sec) | | |
| B | V | R | B | V | R |
| 23.4 | 23.0 | 23.0 | 25.4 | 25 | 25 |
| | | | | | |
| 0.41 Arcsec pixels, S/N=5 | | | | | |
| 1 night (100 sec) | | | 1 year (6000 sec) | | |
| B | V | R | B | V | R |
| 23.2 | 22.8 | 22.8 | 25.2 | 24.8 | 24.8 |

### 2.2 Regions of the sky observed and object counts

Figure 1: It converts equatorial to galactic coordinates

The telescope scans a strip of constant declination equal to the latitude of the observatory. Figure 1 converts equatorial into galactic coordinates and is useful to examine the regions of sky sampled by a zenith telescope in a given site. As the earth rotates and the seasons change, the telescope scans a strip of constant declination moving in and out of the galactic plane. It must be noted that at the latitudes of the best sites (+30, - 30 and + 20 degrees) the telescope samples interesting regions of the sky. We see that at +30 degrees, it goes through the galactic pole. At - 30 degrees, it goes through the south galactic pole and also into the bulge and the center of the galaxy. The strip of sky observed from a Chilean site is highlighted in the figure. The width of the strip of sky observed by the CCD is given by

$$S = 0.206 \, n' \, w \, /f, \tag{2}$$

where S is expressed in arcseconds and n' is the number of pixels in the direction perpendicular to the scan. With a 2048X2048 CCD having 0.6 arcseconds pixels, the strip of sky is 20 arcminutes wide. Table 2 shows the areas of the strips of sky covered by a corrector correcting a 1 square degree field as well as for different CCD sizes and image scales.

Table 2
Areas of sky observed from a Chilean site
===========================================================
(square degrees)

| Field | Total | ExtraGalactic |
|---|---|---|
| 1 degree | 312 | 156 |
| 2048*0.62 arcsec | 110 | 55 |
| 2048*0.41 arcsec | 73 | 36 |
| 3072*0.62 arcsec | 165 | 83 |
| 3072*0.41 arcsec | 109 | 55 |

Figure 2 is useful, to estimate the number of objects that one can observe in a strip of sky. It gives the number counts /square degree of quasars (from Hartwick , & Schade 1990) and galaxies (from Lilly, Cowie, & Gardner 1991 and Metcalfe, Shanks, Fong, & Jones 1991), brighter than a given B magnitude, at the galactic poles. Table 3 gives the integrated counts a B<22 and B<24. Let us only consider the strip of "extragalactic" sky having galactic latitude > 30 degrees and assume that the optical corrector of the telescope yields good images over a 1 degree field, well within the performance of existing corrector designs. We would observe 100 to 200 square degrees of '' extragalactic sky'', depending on the latitude of the observatory , and, at a latitude of 30 degrees, roughly $3 \, 10^6$ galaxies and 50,000 quasars with B< 24. For comparison, the total number of quasars in the Hewitt, & Burbidge (1993) catalog contains 7,000 objects gathered in 30 years, but is essentially useless for statistical studies since the objects were identified from a variety of search techniques having poorly quantifiable selection effects.

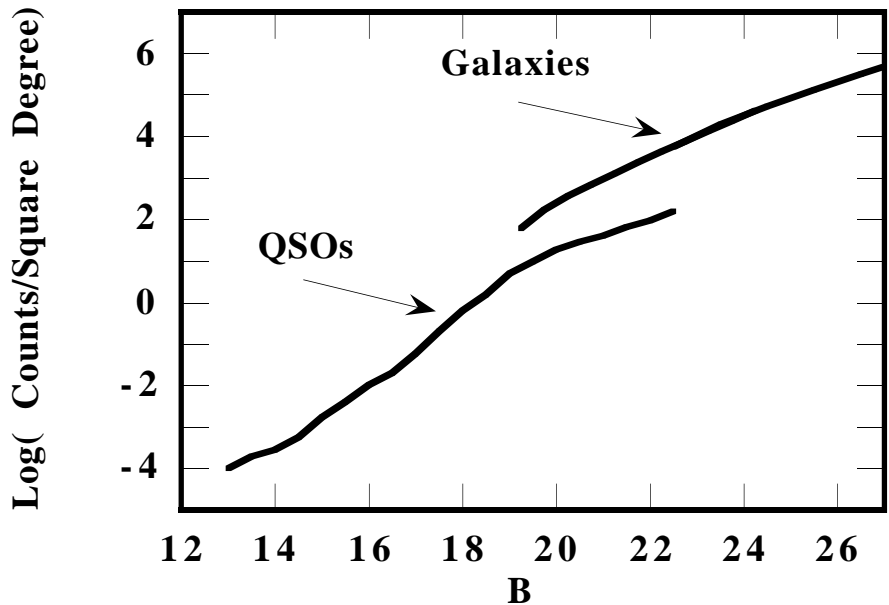

Figure 2: Number counts of quasars and galaxies /square degrees, brighter than a given blue magnitude, at the galactic poles.

Table 3

Integrated object counts
1 degree-wide strip of "extragalactic sky "(bII>30°)
===========================================================

|  | B<22 | B<24 |
|---|---|---|
| Galaxies | 200,000 | 3,000,000 |
| Quasars | 10,000 | 50,000 |

## 3. A WIDE BAND PHOTOMETRIC SURVEY: TARGETS FOR THE VLT

An LMT photometric survey would be quite useful to provide targets to observe with the coming generation of large telescopes (e.g. the VLT). LMTs promise major advances for deep surveys of the sky and, in particular, cosmological studies. Cosmological objects are faint and, furthermore, cosmological studies tend to be statistical in nature; hence need a large number of objects; and therefore the need for considerable observing time on large telescopes. The outstanding limitation of LMTs, that they can only observe near the zenith, is not a serious handicap for cosmological surveys. Let us estimate the number and redshift distributions of galaxies and clusters of galaxies observable with a LMT.

## 3.1 Galaxies

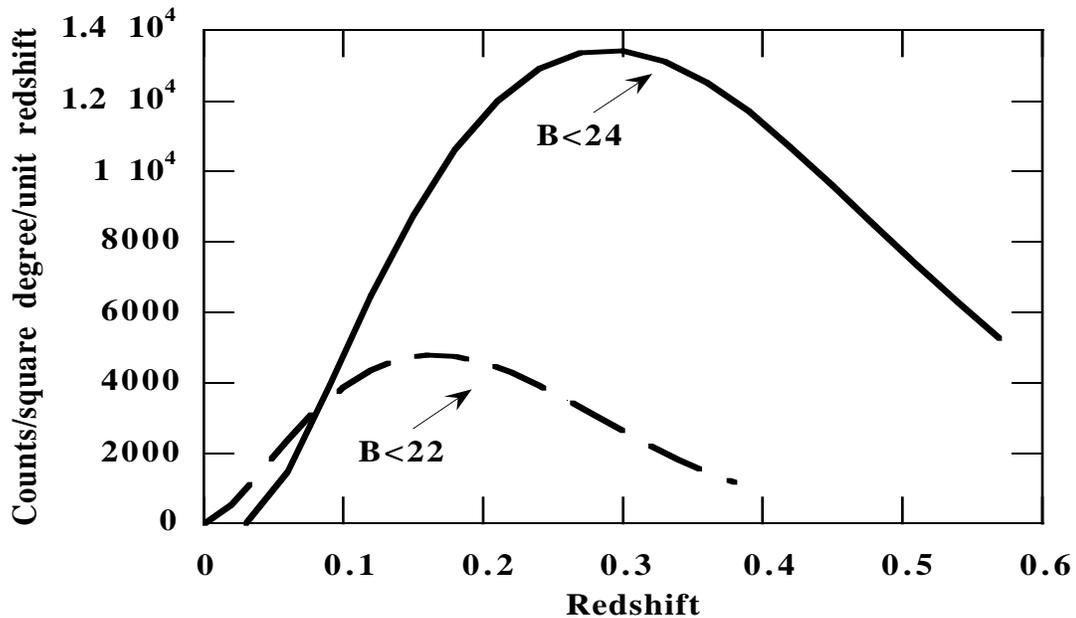

Figure 3: Galaxy redshift distributions expected for surveys reaching 22nd and 24th blue magnitudes.

Table 3 shows that a one-degree strip of sky would access 3,000,000 galaxies to B<24. To see how these observations would sample the Universe, I have computed the redshift distribution expected from a survey having lower limiting magnitude $m_0$ and upper limiting magnitude $m_1$ from the usual cosmological integral

$$\frac{dN}{dz} = d\Omega \int_{m0}^{m1} \Phi(M) \frac{dV}{dz} dm \quad , \quad (3)$$

where $\Phi(M)$ is the differential luminosity function (per unit magnitude) of galaxies, $M(H_0, q_0, z, m)$ the absolute magnitude, m the apparent magnitude, $d\Omega$ the surface area element and $dV(H_0, q_0, z)$ the cosmological volume element/$d\Omega$, $H_0$ the present epoch Hubble constant, $q_0$ the deceleration parameter and z the redshift. I use $H_0 = 100$ km/sec/Mpc and $q_0$=1/2. The models have been computed with the mix of galaxy types and the K-corrections described in Shanks, Stevenson, Fong, & MacGillivray (1984) with the difference that the parameters of the Schechter luminosity function are derived from the CfA survey (de Lapparent, Geller, & Huchra, 1989) . This is an oversimplification since the luminosity function varies with morphology and the Shechter function is an average over all Hubble types. Furthermore, the luminosity function evolves and I neglected evolution, a reasonable but not perfect assumption for the redshift depths involved. The uncertainties brought by these assumptions are tolerable for our purpose since we are only interested in an estimate of the redshift space sampled, rather than detailed modeling. In any event, more sophisticated simulations would rest on assumptions without solid empirical backing. Figure 3 shows the redshift distributions expected for surveys reaching 22nd and 24th blue magnitudes.

The huge database given by nearly 200 square degrees of images could be used for a multitude of projects. Basically one could use the data for the same kind of research that can be done with a Schmidt telescope but with accurate magnitudes and variability.

## 3.2 Clusters of galaxies

We can estimate the number and redshift distributions of clusters of galaxies observed in the area of sky A of the survey from

$$N = n(> M) A \int_{0.1}^{z} \frac{dV}{dz} dz , \qquad (4)$$

where $n(>M)$ is the space density of clusters of galaxies having mass greater than $M$. We use the mass function of clusters of galaxies $n(>M)$ determined by Bahcall and Cen (1993). Table 4 gives the number of clusters observable in the degree-wide strip of extragalactic sky as function of z and richness class. We can see a significant number of clusters for which, for example, we could determine the distribution of mass via lensing of background galaxies, although we will loose part of the cluster for those at the edges of the field. Figure 3 shows that there will be a sufficiently large number of images of galaxies with z > z(cluster) to be able to do the mapping. At z=0.5, the blue magnitude of the third ranked galaxy B3 ~ 23 so that about 30 galaxies should be identifiable to B3+2.0 for R=0 clusters with z<0.5, and more for richer clusters. Bahcall and Cen (1993) give the relations between $n(>M)$, richness class, and number of galaxies in a cluster with B<B3+2.0.

Table 4
Number counts of clusters of galaxies in the degree-wide strip of extragalactic sky as function of z and richness class

==========================================================

| R | 0.1<z<0.3 | 0.1<z<0.4 | 0.1<z<0.5 |
|---|-----------|-----------|-----------|
| 0 | 91 | 187 | 315 |
| 1 | 40 | 82 | 139 |
| 2 | 8 | 16 | 28 |
| 3 | 1 | 2 | 3 |

## 4. VARIABILITY SURVEY AND A SEARCH FOR SUPERNOVAE

Because a LMT observes the same regions of sky night after night, variability "comes for free"; hence variability studies are an obvious application for LMTs.

## 4.1 Type Ia Supernovae

Type Ia supernovae are particularly interesting since they appear to be good standard candles (Branch and Tammann 1992). Taking MB(tmax)= -18.86 (Vaughan et al. 1995) one can readily compute the $m_R$ magnitude at maximum light of type Ia SNs $m_R(z)$ provided one knows the cosmological model and the K correction. We shall assume $H_0 = 75$ and $q_0 = 1/2$ and negligible cosmological constant. Perlmutter et al. ( 1995) have computed the Kcorrection for a SN Ia at z = 0.458 obtaining $K_{BR}$= -0.7. We shall assume that $K_{BR}(z) = K_{BR}(z=0.457)*z/0.457$ and accept some uncertainty due to the Kcorrection, the uncertainty being negligible at z= 0.45 and increasing as one gets away from that redshift. We shall neglect galactic absorption in our galaxy (it is small in R) and the host galaxy. We also neglect the contribution from light from the host galaxies, an assumption is not as bad as it seems since the surface brightness of galaxies is small compared to the sky background, which thus contributes most of the noise. We shall use a peak luminosity $M_B = -18.6$.

Table 5
Apparent R magnitude of a type Ia supernova at maximum light as function of redshift ($H_0 = 75$ and $q_0 = 1/2$)

===========================================================

| z | $m_R$ |
|-----|------|
| 0.6 | 22.4 |
| 0.5 | 22.1 |
| 0.4 | 21.7 |
| 0.3 | 21.2 |
| 0.2 | 20.5 |
| 0.1 | 19.0 |

The light curves of supernovae have time scales long enough that one could coadd data from several nights. This is particularly true for high z objects because of the 1+z time dilation factor. For example, adding 6 nights of data would increase the R limiting magnitude in Table 1 to 24 . Table 5 shows that one can observe the SNs at maximum light at z=0.6. Decent light curves could be obtained for z < 0.4.

## 4.2 Predicted Type Ia supernova rates.

Let us limit ourselves to host galaxies having B<23, a magnitude at which galaxy redshifts can readily be measured with large steerable telescopes. Using the models described by Borra (1995) for B < 23, we see that , in dz=0.1 bins and 1 square degree of sky, there are about 100 galaxies at z=0.1 250 at z=0.2, 400 at z = 0.3 and z= 0.4 and 150 at z=0.5 . Table 2 shows that a single 2048X2048 CCD would cover an extragalactic strip of sky having bII > 30 degrees with a surface from 37 to 56 square degrees so that we would observe nearly 100,000 galaxies in these z ranges. The SN rates depend on the Hubble type of the host galaxy and have been estimated by Van den Bergh and Tammann (1991). For our purposes we can assume a rate of 1 supernova/galaxy/100 years. The survey should therefore find of the order of 1,000 type Ia supernova/year, depending on the actual mixture of Hubble types. This appears to be quite a respectable number of objects. The uncertainty in the rate is about a factor of 2 depending mostly on the mix of galaxy actually observed. It must of course be noted that estimates tend be optimistic and that ours is probably so as well. Using an LMT to find and follow supernovae would save precious time on oversubscribed conventional telescopes that could be better used for spectroscopic follow-up.

### 4.3 Cosmology with type Ia supernovae.

Branch and Tammann (1992) have reviewed the use of type Ia supernovae as standard candles and consider several cosmological research topics. Besides the standard cosmological tests involving the determinations of $H_0$ and $q_0$, the distances obtained from SNs could be used to determine peculiar velocities and streaming motions of the host galaxies. They also mention two tests to determine the nature of the redshift: the surface brightness and the time dilation tests. Note that all of those need follow up spectroscopy with a large telescope. A sample of 1000 supernova/year would allow one to study the characteristics of the sample such as extinction and light-curve dependent luminosity effects. The rates of supernovae, of all types, can be used to infer the rate of massive star formation at z<0.5. This is a time when a surprisingly large number of blue galaxies has been observed, implying a very large rate of star formation.

### 4.4 Other uses of the data

The survey would give an unprecedented sample of variable stars and extragalactic objects. Among several uses:

- Variability of QSOs: Most QSOs have variable luminosities and variability is an efficient way to find them. There should be about 50,000 QSOs with B< 24 in a one-degree strip of sky. One would discover a substantial fraction of them and gather information on their variabilities. QSOs can be identified by their peculiar B-R colors so that variability combined with large B-R colors would allow us to identify them unambiguously.

- Detached double line binaries: Among all types of variable stars, detached double line binaries are particularly interesting since these objects give information on basic stellar parameters (e.g. masses and radii) and, furthermore, are surprisingly good primary distance indicators. They are rare, hence one needs to obtain well determined light curves for a very large number of objects.

- Microlensed galactic objects: Predicting the expected yearly number of Machos requires detailed modeling and is beyond the scope of this work. We can however make very approximate estimates using predicted rates for bulge stars (Griest et al. 1991), leading to an estimated rate at $10^{-7}<G<10^{-6}$. An estimate of the number of stars one will observe in the R band can be obtained from the table of star number densities in the V band given by Zombeck (1990). To V= 21, the number of stars per square degree varies between 200,000 in the galactic plane to 60,000 at a galactic latitude of 20 degrees, a region of sky that can be monitored for half a year and an area of 50 square degrees. The survey would thus observe over 5 million stars in half a year. We would therefore expected from 0.5 to 5 events per year multiplied by an efficiency factor < 0.5. This rate appears low, however note that the predicted rate for the MACHO project observations of bulge stars was also low (<1.1 years) but a significantly higher rate has been found.

### 5. CONCLUSION

The principal advantage of the zenith LMT comes from low cost so that a large telescope can be dedicated to a narrowly focused project and a huge quantity of data can be generated. The low cost of LMTs is an asset for any such project that needs a dedicated medium to large telescope and is not hampered by the limited field of regard.

Fixed LMTs have a major limitation: their reduced sky coverage. However, recent work on innovative correctors (this workshop) indicates that fixed telescopes should be able to access surprisingly large regions of sky (as large as 45X360 degrees wide strips). Estimates of object counts show that, even with the field of view given by conventional correctors, there are large numbers of galaxies, clusters of galaxies and QSOs.

We must carefully distinguish between what can be done now and what might be done later. Fields of views are small with classical glass correctors, so that it is only possible to access a few hundred square degrees of sky with a single LMT; although LMTs located at different terrestrial latitudes will observe different regions of the sky. Tracking has only been demonstrated with the TDI technique so that LMTs can presently only be used as imagers to carry out surveys with filters. Using interference filters one can obtain low resolution spectrophotometry. The type of research that can be done is therefore the type of research that can be done with an imaging survey. Variability studies are obvious applications of LMTs. Among many projects of current interest, a 4-m class LMT would be very competitive for a supernova search.

Considering the present state of the technology, a 4-m class LMT is a low-risk enterprise. We have therefore, right now, a window of opportunity in which one can do what has never been done before in Astronomy: dedicate a 4-m class telescope to a full time project.

If right now it is prudent to consider only a 4-m LMT, in the next few years we can certainly foresee 6 to 8-m LMTs. Given their low costs we can then envision arrays of large LMTs. Given that a 6-m to 8-m fully instrumented LMT can probably be built for the order of 1 million $, one can envision an array of 36 6-m and an array of 64 8-m LMTs for filled collecting areas equivalent respectively to 36-meter and 64-m telescopes at costs comparable to those of present large steerable telescopes.

One can certainly make a wish list of exciting projects feasible with a 64-m telescope; but, arguably, the more interesting contribution of LMTs may come from serendipity. The history of Astronomy tells us that whenever radically new instruments were devised (e.g. radio-telescopes) some totally unexpected major discoveries were made (e.g. quasars, pulsars). Optical astronomy is the oldest of sciences and the telescope was invented centuries ago; however this will be the first time that massive quantities of data will be generated from large telescopes, and computing power is available to analyze them.

## 6. ACKNOWLEDGMENTS

This research has been supported by Natural Sciences and Engineering Research Council of Canada and Formation des Chercheurs et Aide à la Recherche grants. A very large number of people have contributed to the success of Liquid Mirrors. Some of them, but not all, are named in the reference list. I thank them all.